\documentclass{emulateapj}
\usepackage{color}

\slugcomment{draft v2}

\shorttitle{Comprehensive analysis of BH gaps}
\shortauthors{Kisaka et al.}

\begin{document}


\title{Comprehensive analysis of magnetospheric gaps around Kerr black holes using 1D GRPIC simulations}


\author{Shota Kisaka\altaffilmark{1,2}}
\email{kisaka@astr.tohoku.ac.jp}
\author{Amir Levinson\altaffilmark{3}}
\author{Kenji Toma\altaffilmark{1,2}}


\altaffiltext{1}{Frontier Research Institute for Interdisciplinary Sciences, Tohoku University, Sendai, 980-8578, Japan}
\altaffiltext{2}{Astronomical Institute, Graduate School of Science, Tohoku University, Sendai, 980-8578, Japan}
\altaffiltext{3}{School of Physics \& Astronomy, Tel Aviv University, Tel Aviv 69978, Israel}


\begin{abstract}
Spark gaps are likely the source of plasma in active black hole (BH) magnetospheres.  
In this paper, we present results of 1D general-relativistic particle-in-cell simulations of a starved BH magnetosphere with a realistic treatment of inverse Compton scattering and pair production, for a broad range of conditions, run times longer than in previous studies, and different setups.
We find that following the initial discharge,  the system undergoes gradual evolution over
prolonged time until either, restoring the vacuum state or reaching a state of quasi-periodic oscillations, depending on the spectral shape and luminosity of the ambient radiation field.  
The oscillations occur near the null charge surface in cases where the global magnetospheric 
current is in the direction defined by the product of the asymptotic Goldreich-Julian charge density and the radial velocity, while they occur near the boundary of the simulation box when it is the opposite direction (return current). Their
amplitude and the resultant luminosity of TeV photons emitted from
the gap depend sensitively on the conditions;
for the cases studied here the ratio of TeV luminosity to the Blandford-Znajek power ranges from 
$10^{-5}$ to $10^{-2}$, suggesting that strong flares may be generated by moderate changes in 
disk emission.  We also examined the dependence of the solution on the initial number of particles per cell (PPC) and found convergence for PPC of about 50 for the cases examined.  At lower PPC values the pair multiplicity is found to be artificially high, affecting the solution considerably. 
\end{abstract}





\section{Introduction}
\label{sec:introduction}

A question of considerable interest in the theory of force-free outflows from rotating black holes is
the nature of the plasma production mechanism.   Activation of black hole (BH) outflows requires continuous 
injection of plasma at a high enough rate, in order to maintain the density everywhere in the magnetosphere above
the Goldreich-Julian (GJ) value at any time. Since the magnetic field lines forming the outflow penetrate the horizon, and since 
cross-field diffusion of protons is extremely slow, plasma supply from the accretion flow seems unlikely to  accommodate these requirements. Direct injection of electron-positron pairs in the magnetosphere, via annihilation 
of MeV photons emanating from matter surrounding the BH (either the disk or its putative corona) 
may ensue provided it is hot enough, as anticipated in low luminosity AGNs like M87.  
But even then the plasma injection rate might be too low to screen out the magnetosphere 
(\citealt{2011ApJ...730..123L,2016ApJ...818...50H}; but see also \citealt{2020arXiv200313173K}). 

In the absence of sufficient plasma production the magnetosphere becomes charge starved, resulting in the formation of  gaps that can give rise to copious  pair creation through the interaction of particles accelerated in the gap and soft photons emitted by the accretion flow.  Nearly complete screening (i.e., sufficient to maintain the system active) occurs when the pair production opacity contributed by the soft radiation field is sufficiently high.  The density in that case is regulated by a balance between the pair creation rate 
and plasma losses through the inner and outer light surfaces.   This regulation mechanism that, in turn, 
controls the gap dynamics is highly nonlinear and depends very sensitively on the soft photon spectrum
and other details. Analytic models of static gaps around a Kerr BH have been developed \citep{1992SvA....36..642B,1998ApJ...497..563H,2000PhRvL..85..912L,2016ApJ...818...50H,2016ApJ...833..142H,2017PhRvD..96l3006L,2017ApJ...845...77H,2018PhRvD..98f3016F,2020ApJ...895...99K}.  While they provide important insight into the behaviour of these systems, in practice such gaps are expected to be highly intermittent  \citep{2017PhRvD..96l3006L} and, therefore, dynamical models are needed to investigate their properties. 

General relativistic particle-in-cell  (GRPIC) simulations of 1D spark gaps, that incorporate 
radiation processes (Compton scattering, pair creation and 
curvature emission) using  Monte-Carlo methods, have been reported recently in the literature \citep{2018A&A...616A.184L,2020ApJ...895..121C}. 
Such simulations treat the gap as a local disturbance in the global magnetosphere, thereby ignoring  the nonlinear coupling between the gap and the magnetosphere.   On the other hand, they can properly resolve the smallest scales
(the skin depth) and allow relatively long runs to track the long-term evolution of the system.   
 Self-consistent calculations of the global structure and dynamics of the magnetosphere can only be 
 performed using 2D plasma simulations, which are far more demanding.  
Unfortunately, resolving realistic kinetic scales in such simulations is practically infeasible given 
current computing power. To overcome this issue, some rescaling scheme must be invoked \citep{2019PhRvL.122c5101P}.  
This is particularly tricky when radiation processes are included, as they introduce additional scales that 
depend on the conditions in the starved magnetospheric regions in a nonlinear manner.  
The first attempt to perform global 2D simulations that incorporate realistic radiation
processes is described in \citet{2020PhRvL.124n5101C}.  The results indicate that the Blandford-Znajek (BZ) process \citep{1977MNRAS.179..433B} can be 
activated and maintained (at least over the simulation time) if sufficiently intense soft radiation field is present.  However, it is yet unclear how the solution depends on the properties of the ambient radiation field, as well as on subtle numerical issues.  As we shall demonstrate in this paper, in addition to the rapid oscillations that  commence early on, the system also undergoes gradual evolution over a rather long time (tens to hundreds $r_g/c$, depending on the conditions), until converging to a final state.  While global 2D simulations 
are ultimately needed to model the full evolution of the magnetosphere, some 
guidance may still be provided by 1D models.

In this paper we present a comprehensive investigation of 1D gap models using GRPIC simulations similar to those performed by \citet{2018A&A...616A.184L}. A brief summary of the model and the numerical method is described in Sections \ref{sec:estimate} and \ref{sec:setup}. We find (Section \ref{sec:results}) that in most cases, after sufficiently long time the gap dynamics becomes cyclic, as also found in \citet{2020ApJ...895..121C}.  However, the behaviour of the gap depends on the sign of the global electric current; while in the case where the sign is the same as that of the current of outflowing plasma particles with the GJ charge density at the large distance where the frame dragging effect is negligible,
the large amplitude oscillations occur the null surface, in the opposite case the oscillations occur at at the boundary of the simulation box.  The amplitude of the oscillations and the resultant TeV emission depend quite sensitively on the spectrum of the external radiation field; 
for the cases studied here, the luminosity of TeV photons escaping the simulation box was found to range from  $10^{-5}$ to $10^{-2} L_{\rm BZ}$, where $L_{\rm BZ}$ is the BZ luminosity (see Eq. \ref{eq:L_BZ} below).  We also examined how the behaviour of the solutions depends on the initial number of particles per cell and found convergence only when it exceeded about 50. Further discussions are given in section \ref{sec:discussion}, and we conclude in section \ref{sec:conclusion}.

\section{A brief summary of the gap model}
\label{sec:estimate}
A detailed description of the 1D gap model and its assumptions is given in \citet{2018A&A...616A.184L}.  In short, it computes
the evolution of the longitudinal electric field, $E_{\parallel}={\bf E}\cdot{\bf B}/B$, as well as 
the resultant dynamics of accelerated pairs and the various radiation processes, 
in Kerr spacetime, for a given magnetic field topology, 
ignoring the feedback of the gap activity on the global magnetosphere.   The tacit assumption is that
the gap constitutes a small disturbance in the magnetosphere that merely controls the local plasma production, but 
doesn't affect global properties. The coupling between the gap and the global magnetosphere
enters through the global electric current $J_0$ assumed to flow through the gap, 
which is treated in the model as a free input parameter \citep{LM05,2018A&A...616A.184L}. 
For the calculations presented below we adopt a split monopole magnetic field topology. 

The GRPIC code implements Monte-Carlo methods to compute gamma-ray emission and pair-production through 
the interaction of pairs and gamma rays with soft photons emitted by the accretion flow (for details see appendix B in \citealt{2018A&A...616A.184L}).   
The inverse-Compton (IC) gamma rays are treated as a third species of particles (in addition to electrons and positrons) in the code. 
Curvature losses are also included in the calculations, with a power given by
\begin{eqnarray}\label{eq:curvature_power}
P_{\rm cur}=\frac{2}{3}\frac{e^2\gamma^4v^4}{R_{\rm c}^2c^3}
\end{eqnarray}
in the zero angular momentum observer frame, and  with $R_{\rm c}=r_g$ adopted in all runs, where $r_g=GM/c^2$ is the gravitational radius of a BH of mass $M$.
As will be shown below, in certain circumstances curvature emission can dominate 
the total luminosity radiated from the gap.  However, since the characteristic energy of curvature photons is  much lower than that of IC photons  their contribution to pair creation is ignored in order to (considerably) save computing time.  The validity of this assumption is discussed below.   

The intensity of the soft radiation field emitted by the putative accretion flow, that
serves as the source of IC and pair-production opacity in the simulations,
is taken, for simplicity, to be a power law with index $p$ and minimum cutoff 
energy $\epsilon_{\min}$ (that in practice corresponds to the peak of spectral energy distribution):
\begin{eqnarray}
I_{\rm s}=I_0\left(\frac{\epsilon}{\epsilon_{\min}}\right)^{-p}, ~~\epsilon_{\min}<\epsilon<\epsilon_{\max}.
\label{eq:I_0}
\end{eqnarray}
The opacity can be conveniently expressed in terms of a fiducial optical depth $\tau_0$, 
defined as  \citep{2018A&A...616A.184L}
\begin{eqnarray}
\tau_0=\frac{4\pi r_g\sigma_{\rm T}I_0}{hc},
\end{eqnarray}
here $\sigma_T$ and $h$ are the Thomson cross section and Planck constant, respectively.
 The minimum energy $\epsilon_{\min}$, index $p$ and optical depth  $\tau_0$ 
are given as input parameters in addition to global magnetospheric current $J_0$, henceforth
normalized by  $\Omega_H B_H r_H^2(1+a_*^2)\cos\theta/2\pi$, where $B_H=B(r_H)$ is the strength of the  magnetic field on the horizon, at $r=r_H$, $\theta$ is the inclination angle of the magnetic field line, and $\Omega_H$, $a_*$ are the 
angular velocity and dimensionless spin parameter of the BH, respectively (see Appendix for
details). From this definition, the current carried by the outflowing plasma particles with GJ charge density at large distance is $J_0=-1$ (referred as GJ current density), and the return current corresponds to $J_0>0$ (referred as anti-GJ current density, \citet{2013MNRAS.429...20T}). 
To study the dependence of the gap dynamics on these parameters 
we performed the set of simulations  listed in Table 1. 
We considered a positive spin of $\Omega_H>0$ and the radial magnetic field $B_r>0$ in the $0 < \theta < \pi/2$ hemisphere. Then, the GJ charge density is $\rho_{\rm GJ}<0$ beyond the null point. In all cases we used 
$B_H/2\pi= 10^3$ G, $\theta=30^\circ$, $a_*=0.9$ and $M=10^9 M_\odot$.

\begin{table*}
 \label{tab:parameter}
 \caption{Simulation Model Parameters. }
 \begin{center}
  \begin{tabular}{crrllrrll}
\hline
 Model & $J_0$ & $\tau_0$ & $\epsilon_{\min}$ & $p$ & PPC & Time & Initial condition  & \\ \hline
       &       &          &                   &     &     & $(r_g/c)$ &         & \\ \hline
 A      & -1    &  10 & $10^{-8}$  & 2   &  45 &   84 & $e^{\pm}$-filled (A1) & \\
        &       &     &            &     &     &  106 & $\gamma$-filled (A2)  & \\
 B      & -1    & 100 & $10^{-9}$  & 2   &  45 &  241 & Model D & $\tau_0=300\rightarrow100$ \\
 C      & -1    &  30 & $10^{-9}$  & 2   &  45 &  115 & Model B & $\tau_0=100\rightarrow30$ \\
 D      & -1    & 300 & $10^{-9}$  & 2   &  45 &  458 & $\gamma$-filled & \\
 E      & -1    & 100 & $10^{-9}$  & 1.5 &  45 &   90 & Model B & $p=2\rightarrow1.5$ \\
 F      & -1    & 100 & $10^{-9}$  & 3   &  45 &   90 & Model B & $p=2\rightarrow3$ \\
 G      & -1    & 100 & $10^{-8}$  & 2   &  45 &   84 & Model B & $\epsilon_{\min}=10^{-9}\rightarrow10^{-8}$ \\
 H      & -1    & 100 & $10^{-10}$ & 2   &  45 &   72 & Model B & $\epsilon_{\min}=10^{-9}\rightarrow10^{-10}$ \\
 I      & -1    & 100 & $10^{-9}$  & 2   &   5 &  729 & Low-$\tau_0$ & $\tau_0=10\rightarrow100$ \\
 J      & -1    & 100 & $10^{-9}$  & 2   &  15 &  262 & Model M & $\tau_0=300\rightarrow100$ \\ 
 K      & -1    & 100 & $10^{-9}$  & 2   & 135 &  154 & Low-$\tau_0$ & $\tau_0=10\rightarrow100$ \\  
 L      & -1    & 300 & $10^{-9}$  & 2   &   5 & 1045 & Model I & $\tau_0=100\rightarrow300$ \\
 M      & -1    & 300 & $10^{-9}$  & 2   &  15 &  290 & $\gamma$-filled &  \\
 N      &  1    & 100 & $10^{-9}$  & 2   &  45 &  238 & Low-$\tau_0$ & $\tau_0=10\rightarrow100$ \\
 O      &  1    &  10 & $10^{-8}$  & 2   &  45 &  560 & $e^{\pm}$-filled & \\ \hline

  \end{tabular}
 \tablecomments{$e^{\pm}$-filled: a state filled with uniformly distributed electrons and positrons with zero initial velocity and the same number density (left upper panel of Fig. \ref{fig:j-1_tau10_snapshot}). $\gamma$-filled: a state filled with a mono-energetic beam of uniformly distributed gamma-ray photons (left lower panel of Fig. \ref{fig:j-1_tau10_snapshot}). Low-$\tau_0$: an electric-field-screened state after the initial discharge (photon-filled initial condition) in $\tau_0=10$. Model B, D, M or I: a temporal quasi-stationary state of one model.}
 \end{center}
\end{table*}

\subsection{A rough estimate of the critical opacity}
\label{sec:ic_criterion}

Here we provide an approximate criterion for gap screening.
Consider the scattering of seed photons of mean energy $\epsilon_1$ by
an accelerated electron or a positron having a  Lorentz factor $\gamma$, and suppose that the scattered photons, the mean energy of which denoted by $\epsilon_{\rm ic}$, subsequently
collide with  other seed photons of mean energy $\epsilon_2$.
Since we only consider the case $p>1$, for which the IC opacity is dominated by photons near the lower cutoff, we take $\epsilon_1=\epsilon_{\min}$.

Now, in order to self-screen the electric field in the simulation box, the following condition should be satisfied:
\begin{eqnarray}\label{sec3:condition}
\tau_{\rm ic}\tau_{\gamma\gamma}\gtrsim1.
\end{eqnarray}
Omitting a logarithmic factor, the normalized optical depth for IC scattering can be approximated as
\begin{eqnarray}\label{sec3:tau_ic}
\frac{\tau_{\rm ic}}{\tau_0}\sim\left\{ \begin{array}{ll}
1 & (\gamma\epsilon_{\min}\lesssim1) \\
(\gamma\epsilon_{\min})^{-1} & (\gamma\epsilon_{\min}\gtrsim1) , \\
\end{array} \right.
\end{eqnarray}
and the energy of scattered photon is
\begin{eqnarray}\label{sec3:epsilon_ic}
\epsilon_{\rm ic}\sim\left\{ \begin{array}{ll}
\gamma^2\epsilon_{\min} & (\gamma\epsilon_{\min}\lesssim1) \\
\gamma & (\gamma\epsilon_{\min}\gtrsim1). \\
\end{array} \right.
\end{eqnarray}
The pair creation cross section can be expressed as (omitting the logarithmic factor again)
\begin{eqnarray}\label{sec3:sigma_gammagamma}
\frac{\sigma_{\gamma\gamma}}{\sigma_{\rm T}}\sim\left\{ \begin{array}{ll}
0 & (\epsilon_{\rm ic}\epsilon_2\lesssim1) \\
0.1(\epsilon_{\rm ic}\epsilon_2)^{-1} & (\epsilon_{\rm ic}\epsilon_2\gtrsim1), \\
\end{array} \right.
\end{eqnarray}
and it has a maximum for $\epsilon_2\sim\max\{\epsilon_{\rm ic}^{-1},\epsilon_{\min}\}$. 
The pair production opacity is given approximately by $\tau_{\gamma\gamma}\simeq \sigma_{\gamma\gamma}n(\epsilon_2) l$, where
$l$ is the gap width.  With $n(\epsilon_2)\simeq (\epsilon_2/\epsilon_{\min})^{-p}$ from Eq. (\ref{eq:I_0}) and $l\simeq r_g$ one obtains:
\begin{eqnarray}\label{sec3:tau_gammagamma}
\frac{\tau_{\gamma\gamma}}{\tau_0}\sim\left\{ \begin{array}{ll}
0.1(\gamma\epsilon_{\min})^{2p} & (\gamma\epsilon_{\min}\lesssim1) \\
0.1(\gamma\epsilon_{\min})^{-1} & (\gamma\epsilon_{\min}\gtrsim1) \\
\end{array} \right.
\end{eqnarray}
From Eqs. (\ref{sec3:condition}), (\ref{sec3:tau_ic}), and (\ref{sec3:tau_gammagamma}),
the required optical depth is
\begin{eqnarray}\label{sec3:tau0}
\tau_0\gtrsim\sqrt{\frac{\tau_0}{\tau_{\rm ic}}\frac{\tau_0}{\tau_{\gamma\gamma}}}\sim\left\{ \begin{array}{ll}
10^{1/2}(\gamma\epsilon_{\min})^{-p} & (\gamma\epsilon_{\min}\lesssim1) \\
10^{1/2}(\gamma\epsilon_{\min}) & (\gamma\epsilon_{\min}\gtrsim1). \\
\end{array} \right.
\end{eqnarray}

Now, the maximum Lorentz factor of accelerated pairs is limited by curvature losses to
\begin{eqnarray}\label{eq:gamma_max}
\gamma_{\max}&=&\left(\frac{E_{\parallel}R_{\rm c}^2}{e}\right)^{1/4} \nonumber \\
&\sim&1.7\times10^{10}\left(\frac{E_{\parallel}}{B}\right)^{1/4}B_3^{1/4}M_9^{1/2}.
\end{eqnarray}
For the cases studied below $\epsilon_{\min}>10^{-10}$, hence $\gamma_{\max}\epsilon_{\min}>1$. 
Substituting $\gamma_{\max}$ into Eq. (\ref{sec3:tau0}) we obtain $\tau_0>30 (300)$ for $\epsilon_{\min}=10^{-9}(10^{-8})$, which corresponds to $\approx0.5~{\rm meV}(5~{\rm meV})$.  This estimate is found to be in reasonable agreement
with the simulations.  Note that the maximum Lorentz factor in Eq. (\ref{eq:gamma_max})
depends on the strength of the gap electric field, which in certain regions is well below $B$.
In the $J_0=+1$ models we find that the critical opacity is smaller by up to an order of
magnitude than the naive estimate obtained for $E_{\parallel}=B$.

\begin{figure*}
 \begin{center}
  \includegraphics[width=160mm]{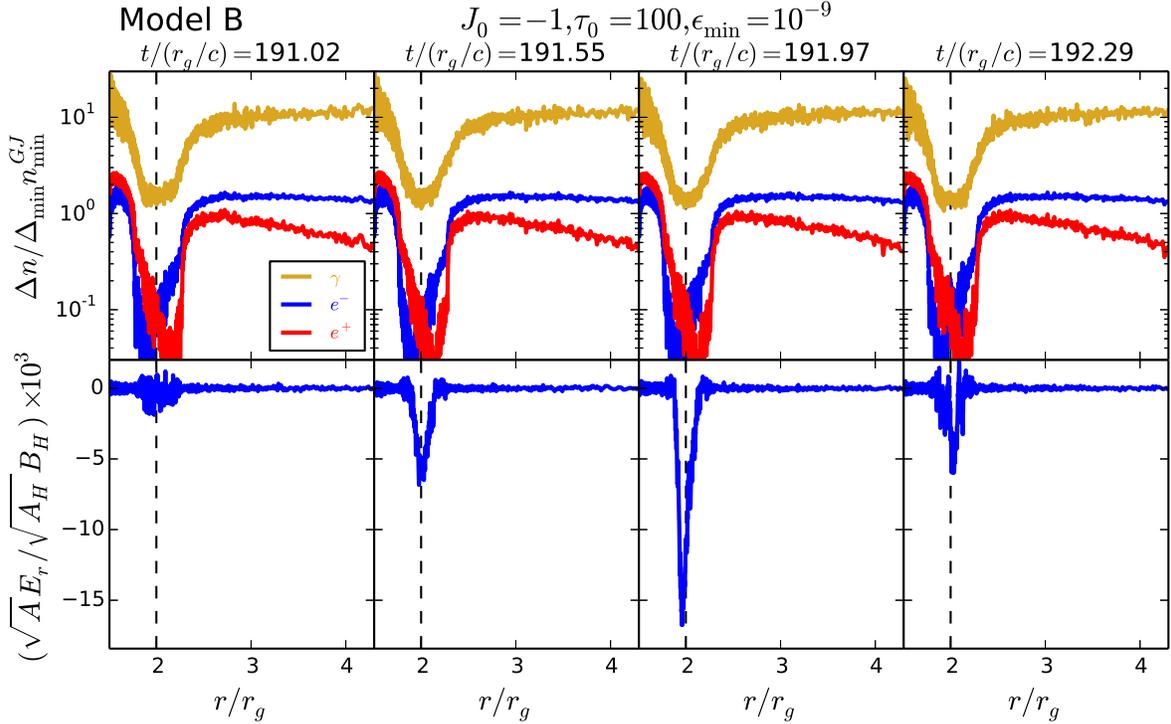}
   \caption{Snapshots of a typical gap cycle from the simulation of model B, between the times indicated above each panel.  The top panels show the evolution of the normalized densities of electrons (blue), positrons (red) and photons (yellow) by $\Delta_{\min}n_{\min}^{\rm GJ}\equiv\Delta(r_{\min})n_{\rm GJ}(r_{\min})$, and the bottom panels show the evolution of the electric field.  The vertical dashed line marks the null charge surface. Note that the normalized densities in the top panel are different than the local pair multiplicities, defined as the number density of electrons and positrons normalized by the local GJ density, 
   since the GJ density is not uniform. The local pair multiplicities are greater than unity everywhere inside the simulation box.}
  \label{fig:j-1_tau100_snapshot}
 \end{center}
\end{figure*}

The above considerations ignore the contribution of curvature photons to pair creation. To assess its relative importance, note that 
the following condition must be satisfied in order that pair production by curvature photons will suffice to screen out the gap:
\begin{eqnarray}\label{eq:condition_cur}
N_{\rm c}\tau_{\gamma\gamma}\gtrsim1,
\end{eqnarray}
where $N_{\rm c}$ is the number of curvature photons emitted by a single 
particle in the gap.  The latter can be estimated as 
\begin{eqnarray}\label{eq:N_c}
N_{\rm c}&\sim&\frac{P_{\rm cur}}{\epsilon_{\rm c}m_{\rm e}c^2}\frac{l}{r_g},
\end{eqnarray}
where $\epsilon_{\rm c}$ is the characteristic energy of the curvature radiation, 
\begin{eqnarray}\label{eq:epsilon_c}
\epsilon_{\rm c}&=&\frac{3}{4\pi}\frac{h}{R_{\rm c}m_{\rm e}c}\gamma^3 \nonumber \\
&\sim&3\times10^5M_9^{-1}\gamma_{10}^3, 
\end{eqnarray}
which corresponds to $\approx200~M_9^{-1}\gamma_{10}^3$ GeV.
In the gap, the number of accelerated particles is $\sim|j_0|/ec\sim n_{\rm  GJ}\sim B\Omega/(2\pi ec)$, where $\Omega=a_{\ast}c/4r_g$. 
In terms of the curvature luminosity $L_{\rm cur}$, the energy loss rate of a single
particle due to curvature emission can be expressed as
\begin{eqnarray}\label{eq:P_cur}
P_{\rm cur}&\sim&\frac{L_{\rm cur}}{4\pi r_g^2ln_{\rm GJ}} \nonumber \\
&\sim&\frac{ec}{8}\left(\frac{L_{\rm cur}}{L_{\rm BZ}}\right)\left(\frac{l}{r_g}\right)^{-1}B,
\end{eqnarray}
where 
\begin{eqnarray}\label{eq:L_BZ}
L_{\rm BZ}=\frac{1}{16}a_{\ast}^2B_{\rm H}^2r_{\rm H}^2
\end{eqnarray}
is the BZ luminosity. From Eqs. (\ref{eq:N_c}) and (\ref{eq:P_cur}), the number of curvature photons is given by
\begin{eqnarray}\label{eq:N_c2}
N_{\rm c}\sim3\times10^7\left(\frac{L_{\rm cur}}{L_{\rm BZ}}\right)B_3M_9^2\gamma_{10}^{-3}
\end{eqnarray}
For the pair-creation optical depth $\tau_{\gamma\gamma}$, the energy of the target photon is $\epsilon_2\sim\epsilon_{\rm c}^{-1}>\epsilon_{\min}$. 
Then, the ratio of the optical depth is 
\begin{eqnarray}
\frac{\tau_{\gamma\gamma}}{\tau_0}\sim\left(\frac{\epsilon_{\rm c}^{-1}}{\epsilon_{\min}}\right)^{-p}\frac{\sigma_{\gamma\gamma}}{\sigma_{\rm T}}\frac{l}{r_g}.
\end{eqnarray}
If the index is $p=2$, the optical depth is
\begin{eqnarray}\label{eq:tau_gammagamma_p2}
\frac{\tau_{\gamma\gamma}}{\tau_0}\sim9\times10^{-9}\left(\frac{l}{r_g}\right)\gamma_{10}^6M_9^{-2}\epsilon_{\min,-9}^2.
\end{eqnarray}
Using Eqs. (\ref{eq:N_c2}) and (\ref{eq:tau_gammagamma_p2}), the criterion for the optical depth (\ref{eq:condition_cur}) is described by
\begin{eqnarray}\label{eq:curvature_condition}
\tau_0\gtrsim4\left(\frac{L_{\rm cur}}{L_{\rm BZ}}\right)^{-1}\left(\frac{l}{r_g}\right)^{-1}B_3^{-1}\gamma_{10}^{-3}\epsilon_{\min,-9}^{-2}.
\end{eqnarray}
Using the simulation results for $L_{\rm cur}/L_{\rm BZ}$, we can check whether the curvature photon pair-creation is important or not.

\section{Numerical setup and method}
\label{sec:setup}
We use the same one-dimensional hybrid Particle-In-Cell/Monte Carlo hybrid code described in \citet{2018A&A...616A.184L}.  It uses the Boyer-Lindquist coordinates (see Appendix), with 
the radial coordinate replaced by the tortoise coordinate $\xi$ to avoid the singularity on the horizon.  
The spatial grid is fixed in time and uniform in $\xi$, ranging from $\xi_{\min}=-3~(r_{\min}/r_g\sim1.5)$ to $\xi_{\max}=-0.3~(r_{\max}/r_g\sim4.3)$. The inner boundary is close to the horizon, $r_H/r_g=1.44$ for $a_{\ast}=0.9$. We check that the results do not change for the case of $\xi_{\min}=-2.5$.  We fix the number of grid cells to be $N_{\xi}=32768$. We check that the grid size is always smaller than the plasma skin depth. The timestep is set by the Courant-Friedrichs-Lewy condition defined at the inner boundary. For the boundary conditions we use open boundaries, that is, the particles, whether charged or neutral, are simply deleted when crossing the boundary on either side. No plasma injection from the outside of the boundaries is assumed. As an indicator of particles per cell (PPC), we use the number of simulation photons per cell for the photon-filled initial condition (see below) with the density 10$n_{\min}^{\rm GJ}$ where $n_{\min}^{\rm GJ}$ is the Goldreich-Julian number density at the inner boundary. The resolution and numerical convergence are checked as described in Section \ref{sec:numerical_effects}.

We use three types of initial conditions; (a) the simulation box is filled with a mono-energetic beam of gamma-ray photons with outward direction uniformly distributed along the $\xi$-grid. The energy and number density of photons are $10^9m_{\rm e}c^2$ and $10n_{\min}^{\rm GJ}$, respectively; (b) the box is filled with electrons and positrons uniformly distributed, with zero initial velocity and the same number density, $0.8n_{\min}^{\rm GJ}$, so that the initial charge and current densities are zero; (c) To save computing time in long runs, we switch from a temporal quasi-stationary state of one model to another model by changing the set of parameters.  In conditions (a) and (b), the initial electric field is so strong that many pairs and photons are produced during the initial discharge as shown in \citet{2018A&A...616A.184L}. In order to avoid this intense discharge and reduce the computational costs, we mainly use condition (c). The initial conditions we use are summarized in Table 1; the rightmost column indicates the parameters of the initial model before switching. The quasi-steady state of condition (c) used in models I, K and N is the electric-field-screened state after the initial discharge (photon-filled initial condition) in the low-$\tau_0$ model ($\tau_0=10$) and prior to re-opening of the gap. We tested models with different initial conditions, and find that the final state is independent of the initial model invoked (e.g., Fig. \ref{fig:j-1_tau10_snapshot}).

To reduce the computational cost in runs with large optical depths ($\tau_0\ge30$), 
we neglect the scattering of particles with energies $\gamma<\gamma_{\rm min,scat}=10^7$. 
The energy of photons scattered by these pairs is $\epsilon_{\rm ic}\sim10^5\gamma_7^2\epsilon_{\min,-9}$, 
much lower than the threshold energy for pair creation, $\sim10^9\epsilon_{\min,-9}$. 
For sufficiently steep spectra ($p\ge2$), the contribution of 
these gamma-ray photons to pair production is minor. 
Note that cooling via IC scattering could still be significant for pairs with $\gamma\sim10^7$ because the cooling length is $l_{\rm ic,cool}\sim r_g\gamma_7^{-1}\epsilon_{\min,-9}^{-1}\tau_{0,2}^{-1}$. 
As a check, we repeated the run for $\tau_0=300$ with a lower threshold of $\gamma_{\rm min,scat}=10^6$ and confirmed that the results remained practically unchanged. 

\section{Results}
\label{sec:results}
This study extends the previous works \citep{2018A&A...616A.184L,2020ApJ...895..121C} in several ways.  First, we performed much longer
simulations to check the prolonged evolution of the system. Indeed, as shown below, we find that the time it takes 
the gap to either restore (if the opacity is to low) or reach a quasi-periodic state is considerably longer than the simulation times in our previous study.   Second, we tested the behaviour of the gap dynamics on the sign of the global current $J_0$. We find a different behaviour for currents with $J_0<0$ and $J_0>0$. 
Third, we have run simulations with larger fiducial optical depths than previously ($\tau_0=30, 100, 300$) and found that although the particle kinetic power and IC luminosity do not significantly depend on $\tau_0$, 
the the luminosity of curvature radiation shows a strong dependence (Section \ref{sec:Negative}). 
Fourth, we checked the dependence of gap dynamics on the spectral shape of the target radiation field (slope $p$ and minimum energy $\epsilon_{\min}$, Section \ref{sec:spectrum_dependence}). Finally,  we investigated the dependence of the solutions on the number of particle per cell (PPC) (Section \ref{sec:numerical_effects}). We find that while the strength and overall dynamics of the gap electric field show only little dependence on the PPC, the pair multiplicity, the pair number density normalized by the GJ density $n/n_{\rm GJ}$, depends strongly on it.  Moreover, we repeated 
the simulation presented in \cite{2018A&A...616A.184L} (Model O in Table 1 with initial PPC of 5) and found that 
when the run time is increased sufficiently their result at the final simulation time (the rightmost panels of their Fig. 1) is not maintained. 
When the simulation was rerun with initial PPC of 45 the gap has ultimately reached a state of quasi-periodic oscillations at $t\sim200~r_g/c$, which is $\sim$10 times longer than the final state in \citet{2018A&A...616A.184L}.  

\subsection{Dependence of gap dynamics on the global current}
\label{sec:current_dependence}
\begin{figure}
 \begin{center}
  \includegraphics[width=80mm]{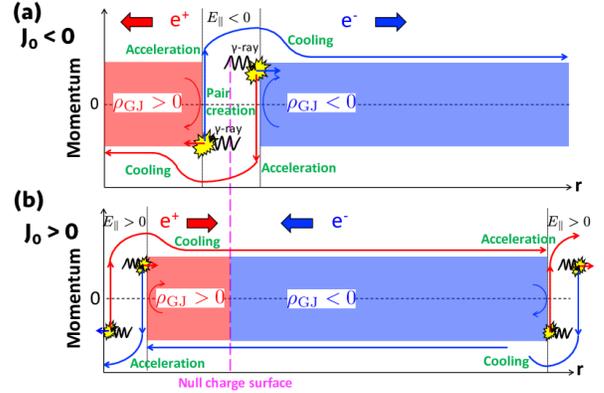}
   \caption{Schematic illustration of the gap structure following the initial discharge phase, for $J_0<0$ (a) and $J_0>0$ (b).  The thick arrows indicate the slow drift (fluid) motion of the plasma constituents. Depending on the sign of the electric field in the reopened gap, plasma moves either away from or towards the null point. This leads to the distinct behaviour of current flows with $J_0>0$ and $J_0<0$ described in Section \ref{sec:current_dependence}. In the reopening gap, particles are accelerated and emit high-energy $\gamma$-ray photons. A fraction of the photons convert to pairs which tend to prevent further growth of the reopened gap. The energy of the accelerated particles significantly decreases outside this gap via curvature radiation. }
  \label{fig:interpretation}
 \end{center}
\end{figure}
\begin{figure}
 \begin{center}
  \includegraphics[width=90mm]{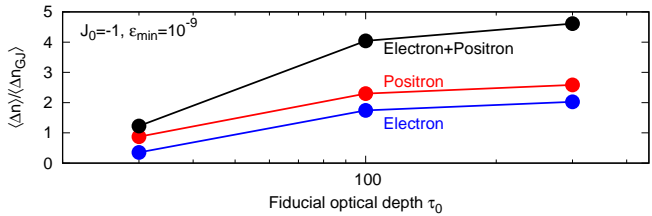}
   \caption{Time-averaged number density of electrons (blue), positrons (red), and the sum of them (black) in the simulation box for models B-D, normalized by the averaged GJ number density in the simulation box, $\langle \Delta n_{\rm GJ}\rangle$, as a function of the fiducial optical depth $\tau_0$.}
  \label{fig:tau_dependence}
 \end{center}
\end{figure}
\begin{figure*}
 \begin{center}
  \includegraphics[width=140mm]{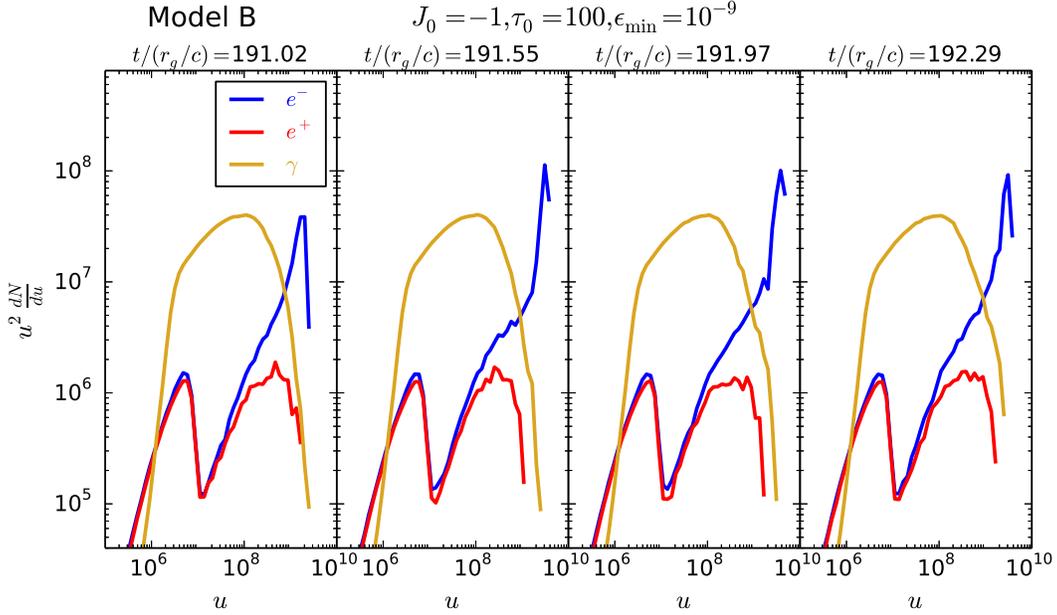}
   \caption{Snapshots of energy spectra of outgoing electron (blue curve), positron (red curve), and scattered photon (yellow curve) in the box from the same simulation presented in Fig. \ref{fig:j-1_tau100_snapshot}.}
  \label{fig:j-1_tau100_spectrum}
 \end{center}
\end{figure*}
As mentioned above, the behaviour of the gap appears to depend on the sign of the global magnetospheric current $J_0$.
To be concrete, for all cases explored here we find that after sufficiently long time from the initial discharge, reopening of the gap starts near the null point when the sign of the current is the same as GJ current ($J_0<0$) and near one of the outer boundaries of simulation box when the current is anti-GJ one ($J_0>0$) \footnote{A polar angle $\theta=30^{\circ}$ would seem to be small for the return current region $(J_0>0)$. However, for example, in 2D PIC simulation results \citep{2019PhRvL.122c5101P}, the boundary of the return current region could extend to the polar angle $\theta\sim40-50^{\circ}$ near the horizon. So, the polar angle we use for the current $J_0>0$ is not unreasonable.}.
In both cases, if the 
fiducial depth $\tau_0$ is below the critical value needed for screening (roughly the value given in Eq. \ref{sec3:tau0}) the charge density gradually declines until all charges escape
from the simulation box and the vacuum state is restored.  On the other hand, if $\tau_0$ is sufficiently large the gap ultimately reaches a state of quasi-periodic oscillations, similar to those found in \citet{2020ApJ...895..121C}. As will be described in greater detail below, when $J_0<0$ the oscillations occur predominantly near the null point (see Fig. \ref{fig:j-1_tau100_snapshot}) whereas 
for $J_0 >0$ they occur near the boundary.\footnote{In all cases rapid plasma oscillations are superposed on the large amplitude cycles.} 

The aforementioned dependence on $J_0$ can be traced to the dynamical equation for the gap electric field $E_r$ (see, e.g., \citealt{2017PhRvD..96l3006L}),
\begin{equation}
\partial_t(\sqrt{A} E_r) =- 4\pi(\Sigma j^r-J_0),\label{eq:dEdt}
\end{equation}
here $A$ and $\Sigma$ are the metric components defined in Eq. (\ref{eq:metric2}) and $j^r$ is the electric current carried by the pairs inside the gap.  The initial condition is set up by solving Gauss equation for a given initial distribution of electrons and positrons in the simulation box.   With our convention ($\rho_{GJ}<0$ beyond the null point) the electric field in the simulation box is initially negative, $E_r<0$. Upon screening by the plasma supplied following the initial discharge episode, the time averaged current, $\langle j^r\rangle$, approaches $J_0$ and $E_r$ undergoes 
small amplitude oscillations around $\langle E_r\rangle=0$.  As time passes plasma starts escaping the simulation box, driving $\langle E_r\rangle$ towards negative values when $J_0<0$ and positive values when $J_0>0$, as can be seen from Eq. (\ref{eq:dEdt}) for $|\langle j^r\rangle| <|J_0|$.  This leads to a drift plasma motion away from the null point in the former case and through 
the null point in the latter case,  and the consequent opening of the gap at these points, as shown schematically in Fig.  \ref{fig:interpretation}. 
If $\tau_0$ is not large enough the vacuum state will ultimately be restored.  Otherwise, a moderate opening of the gap leads to rapid pair creation that suffices to
replenish the plasma lost from the boundaries, and to maintain the gap active through large amplitude cycles.

\subsection{GJ electric current flow}
\label{sec:Negative}
\begin{figure*}
 \begin{center}
  \includegraphics[width=180mm]{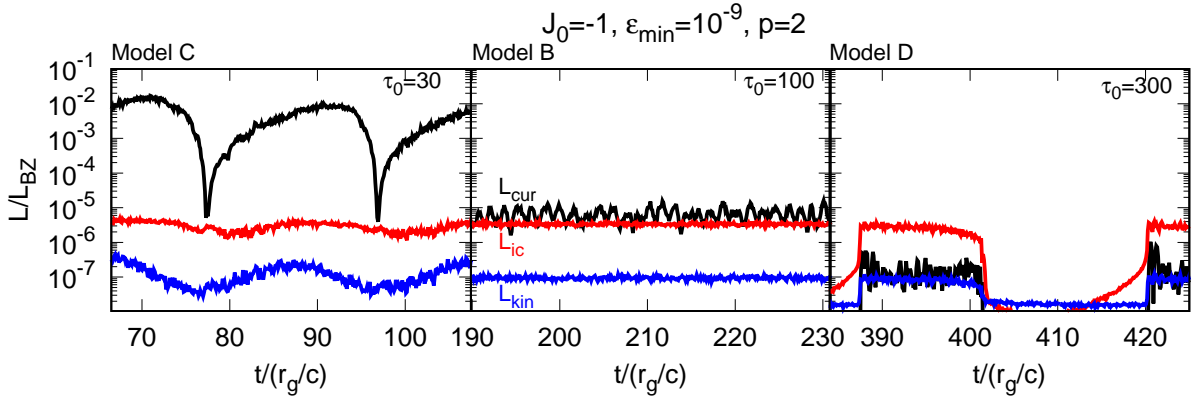}
   \caption{Light curves of curvature radiation (black), IC emission (red), and particle kinetic luminosity (blue) at the outer boundary of the simulation box, from runs with $J_0=-1, \tau_0=30$ (left), $100$ (middle) and $300$ (right), and $\epsilon_{\min}=10^{-9}$. All luminosities are normalized by the BZ power given in Eq. (\ref{eq:L_BZ}).}
  \label{fig:j-1_lightcurve}
 \end{center}
\end{figure*}
We begin by considering the results of simulations with GJ current, $J_0=-1$ (Models $A - M$ in Table 1).  
We find a rough agreement with the analytic criterion for gap screening: $\tau_0\gtrsim30\gamma_{10}\epsilon_{\min,-9}$ (Eq. \ref{sec3:tau0}).
In all models with $\tau_0\ge30$ in Table 1 the gap reaches an active, quasi periodic state following the initial 
discharge phase, independent of initial conditions, whereas in models with $\tau_0=10$  the gap ultimately returned to the vacuum state.

A typical cycle is shown in Fig. \ref{fig:j-1_tau100_snapshot}, where the evolution between the times $t/(r_g/c)=191$ and $192$ is 
displayed for model B. 
This cycle repeats itself over the entire simulation time, with no signs for a gradual change in the overall dynamics. 
As seen, the gap electric field exhibits large amplitude oscillations around the null surface (at $r=2 r_g$) with a period 
of $\sim1.3(r_g/c)$.  Small amplitude, rapid plasma oscillations are superposed on these cycles.  The pair and photon
densities drop by more than an order of magnitude in the region where the gap opens periodically; the photons are produced predominantly outside this region. 
We emphasize that the period of the large amplitude oscillations is  shorter than the light crossing time of the simulation box and is, therefore, 
not prone to numerical effects  (specifically the size of the computational box).
The maximum width of the oscillating gap is $l_{\rm osc}\sim0.2r_g$, 
consistent with the requirement $\tau_{\rm ic}\tau_{\gamma\gamma}\sim10\gamma_{10}^{-2}\epsilon_{\min,-9}^{-2}\tau_{0,2}^2(l_{\rm osc}/r_g)\approx 1$.
Inside this gap,  positrons accelerate in the inward direction and electrons in the outward direction by virtue of the negative sign of the electric field.

A similar behaviour is seen in models C,D.   In general,
the amplitude of the oscillating electric field increases with decreasing $\tau_0$.  
The oscillation period, on the other hand, shows no clear trend;
for $\tau_0=30$ it is nearly 10 times longer than in model B ($\sim 20 r_g/c$), 
and likewise for $\tau_0=300$ it is also longer than in model B. These trends are reflected in the lightcurves shown in Fig. \ref{fig:j-1_lightcurve} (further discussed below).  In all cases, the average pair multiplicity is found to be of order unity, as seen in Fig. \ref{fig:tau_dependence}.

Fig. \ref{fig:j-1_tau100_spectrum} shows the energy spectrum of outgoing electrons, positrons, and IC scattered photons in model B.
It indicates that the maximum energy of outgoing electrons increases as the amplitude of the oscillating electric field increases.  This maximum energy is dictated by a balance between the rate of energy gain by acceleration and the loss rate due to curvature radiation. 
For this reason, the curvature luminosity could exceed the particle kinetic luminosity.  
As seen in Fig. \ref{fig:j-1_tau100_spectrum}, the energy distribution of both electrons and positrons is bimodal. 
Such energy distribution is consistent with results of PIC simulations of  pulsar gaps 
\citep{2010MNRAS.408.2092T,2013MNRAS.429...20T,2016ApJ...829...12K,2018ApJ...855...94P}. 
The low energy component does not significantly contribute to the electric current, but rather adjusts the charge density to the local GJ value. 
The high energy pairs adjust the current density to the global magnetospheric current, so that the number density is $\sim|j_0|/ec$. 

Fig. \ref{fig:j-1_lightcurve} exhibits light curves for models $B-D$. 
The black and red lines delineate the contribution of curvature and IC emissions, respectively.
The luminosities are given as ratios of the BZ power (Eq. \ref{eq:L_BZ}).
As seen, the luminosity of curvature radiation is very sensitive to the fiducial depth $\tau_0$; it varies  from  $\sim10^{-2}L_{\rm BZ}$ for $\tau_0=30$ to about $\sim10^{-6}L_{\rm BZ}$ at $\tau_0=300$.  In contrast, the IC luminosity is  independent of $\tau_0$ and is fixed at $\sim10^{-5}L_{\rm BZ}$, consistent with the value computed in \citet{2018A&A...616A.184L}.  As shown below, the curvature luminosity is also sensitive 
to the spectral shape of the external radiation.  This implies that strong flares can be produced by moderate changes of the disk luminosity.   The characteristic energy of the curvature photons in Model C ($\tau_0=30$), where it completely dominates the output, is of the order of 1 TeV (see Eq. \ref{eq:epsilon_c}). 

Using condition (\ref{eq:curvature_condition}) we estimate the potential contribution of curvature photons to pair creation, which we neglected. In Model C, the luminosity of curvature radiation is $L_{\rm cur}\sim 10^{-2}L_{\rm BZ}$ (see left panel of Fig. \ref{fig:j-1_lightcurve}).  From Eq. (\ref{eq:curvature_condition}) we then find that the optical depth required to screen out the electric field by curvature pair creation is $\tau_0\gtrsim 400$, much larger than actual value, $\tau_0=30$. Consequently, the contribution of curvature photons to pair creation is subdominant in Model C. However, for lower $\tau_0$ cases, such as Model A, a higher ratio of $L_{\rm cur}/L_{\rm BZ}$ is expected, hence pair creation by curvature photons might be important in those cases. This might change, somewhat, the threshold value of $\tau_0$ needed for screening.

\begin{figure*}
 \begin{center}
  \includegraphics[width=140mm]{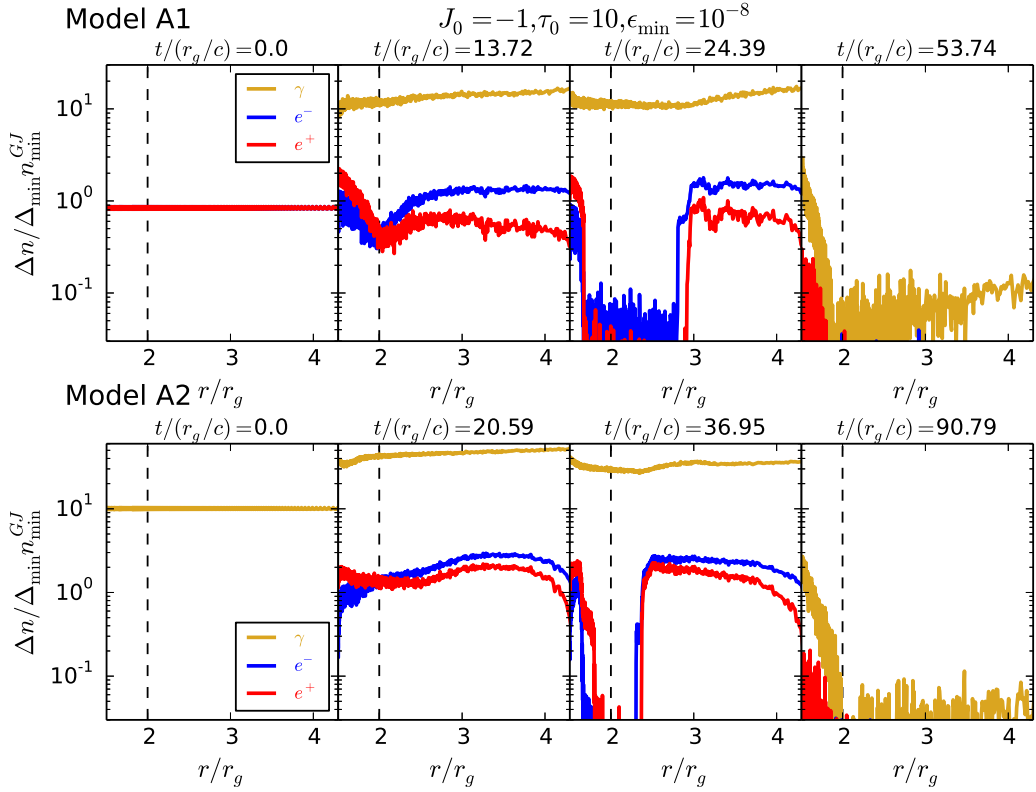}
   \caption{Snapshots form the simulations of the subcritical models $A_1$ (upper panels) and $A_2$ (bottom panels),
   showing the evolution of the pairs and scattered photon densities. 
   The leftmost panels delineate the initial state, at $t=0$. The rightmost panels show the nearly vacuum state reached by the system at the end of the simulation. 
   }
  \label{fig:j-1_tau10_snapshot}
 \end{center}
\end{figure*}
\begin{figure*}
 \begin{center}
  \includegraphics[width=160mm]{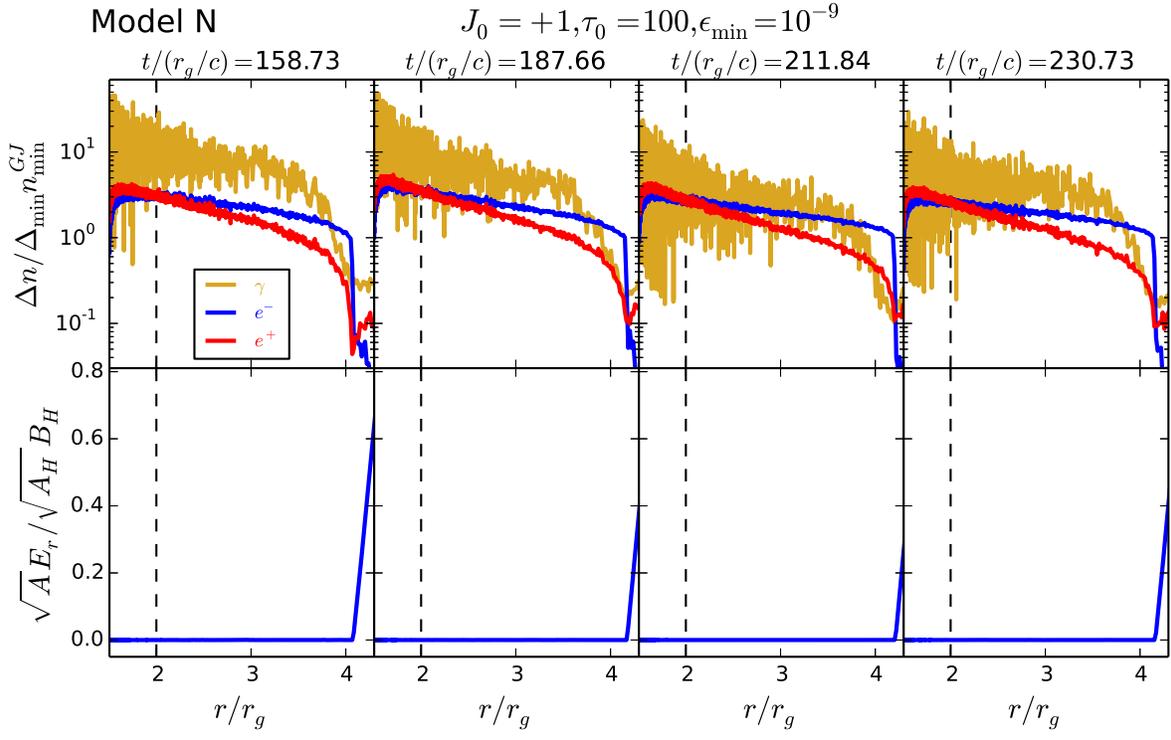}
   \caption{The evolution of the photon and pair densities (upper panel) and electric flux (lower panel) during a typical cycle, similar to Fig. \ref{fig:j-1_tau100_snapshot}, in a simulation with $J_0=+1, \tau_0=100$ and $\epsilon_{\min}=10^{-9}$ (model N). As seen, the cyclic opening of the gap occurs at the outer boundary in this model.}
  \label{fig:j+1_tau100_snapshot}
 \end{center}
\end{figure*}

\begin{figure*}
 \begin{center}
  \includegraphics[width=180mm]{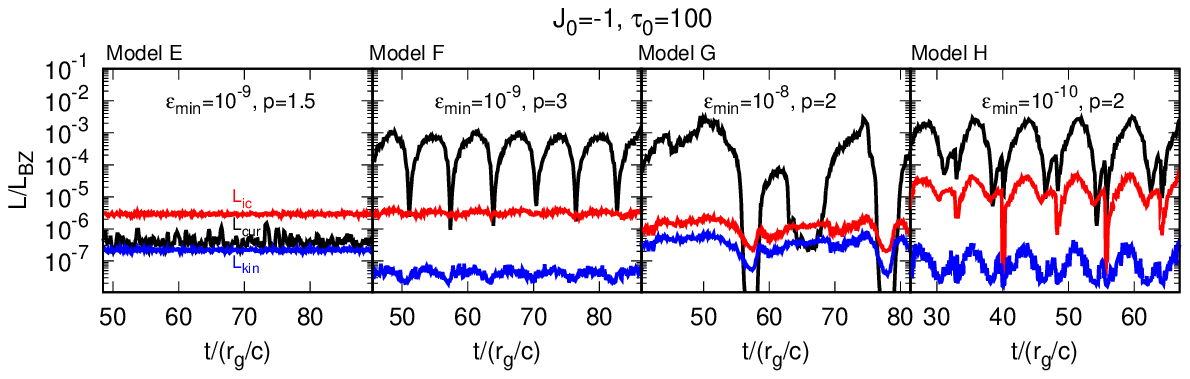}
   \caption{Light curves obtained from the simulations of models E-H, as indicated.
   }
  \label{fig:fig:j-1_tau100_epsilon-8_lightcurve}
 \end{center}
\end{figure*}

We now turn to discuss the sub-critical model (model A).  For this model we explored two different setups (models A1 and A2 in Table 1), distinguished by the initial distribution of pairs and gamma rays (see Fig. \ref{fig:j-1_tau10_snapshot});
in the first one the gamma ray density is initially zero and pairs are uniformly injected in the simulation domain (such that the charge density is zero). In the second one the pair density is initially zero
and photons are uniformly injected.  The subsequent evolution of the density of pairs and gamma rays for both models is shown in 
the upper and lower panels of Fig.  \ref{fig:j-1_tau10_snapshot}.   In both cases the initial discharge produce sufficient charges to screen out 
the gap electric field.  However, as time passes pairs and photons escape the simulation box and the rate at which new pairs are 
 created becomes too low to replenish the lost plasma.  As a consequence, the gap opens near the null surface and gradually grows, as seen in the figure,
 until the simulation box becomes completely devoid of pairs and gamma rays and the vacuum state is restored.

\subsection{Return electric current flow}
\label{sec:Positive}

As explained above, the main difference between the current flows with $J_0>0$ and $J_0 < 0$  is the tendency of the gap to reopen at the outer boundaries in the former case rather than near the null surface.   Depending on the choice of parameters, a gap spontaneously reopens after the initial discharge either at the inner or outer  boundary of the simulation box.  When $\tau_0$ is small (roughly below 10) the vacuum state is ultimately restored, as in the GJ current cases ($J_0=-1$).  At larger values we find quasi-periodic oscillations at the outer boundary, as seen for example  in Fig. \ref{fig:j+1_tau100_snapshot}. We find an active gap even in model O, owing to cooling of positrons outside the oscillating region.  However, for lower $\tau_0$ ($< 10$) the gap 
ultimately returns to the  vacuum state.

We note that the behaviour observed in the return current models might be affected by the limited extent of the computational domain. In practice, particles accelerated by the strong electric field near the boundary will generate pair cascades in the region outside of the simulation box.  This additional injection of charges near the boundary,  which is unaccounted for in our simulations, can lead to partial screening of the electric field in the opened gap.  How this might affect the overall dynamics is unclear at present; we anticipate that the gap might expand sideways. 


\subsection{Dependence on the target spectrum}
\label{sec:spectrum_dependence}
As we now show, the gap dynamics is quite sensitive to the properties of the target, soft 
photon spectrum, owing to Klein-Nishina (KN) and threshold effects.  As a rule of thumb,
when the cutoff energy, $\epsilon_{\min}$, is large all interactions (that is, IC scattering 
and pair production) are in the deep KN regime.  On the other hand, when $\epsilon_{\min}$ is 
small enough and the spectrum is not too flat ($p\ge2$), the number of photons having  energies
near the threshold, $\epsilon_{\rm th} \simeq \langle\gamma\rangle^{-1}$, where $\langle\gamma\rangle$ is the mean Lorentz factor
of the accelerated pairs, is a fraction $\langle\gamma\rangle\epsilon_{\min}$ of the total, and the opacity 
is reduced. Both cases lead to a significant suppression of the pair creation rate. 
When the spectral slope is hard (soft), the pair creation optical depth for gamma-ray photons whose energy is lower than $\epsilon_{\min}^{-1}$ is increased (decreased). 
This can have a dramatic effect on the gap activity.

Examples are shown in Fig. \ref{fig:fig:j-1_tau100_epsilon-8_lightcurve}, where 
lightcurves obtained for $J_0=-1$, $\tau_0=100$ and different values of $p$ and $\epsilon_{\min}$
are displayed (see middle panel in Fig. \ref{fig:j-1_lightcurve} for an additional case). 
These experiments imply that a moderate change in the slope and/or spectral peak of the
ambient radiation field can lead to a dramatic change in the amplitude of the oscillating 
gap electric field, and the consequent curvature luminosity.

\begin{figure}
 \begin{center}
  \includegraphics[width=90mm]{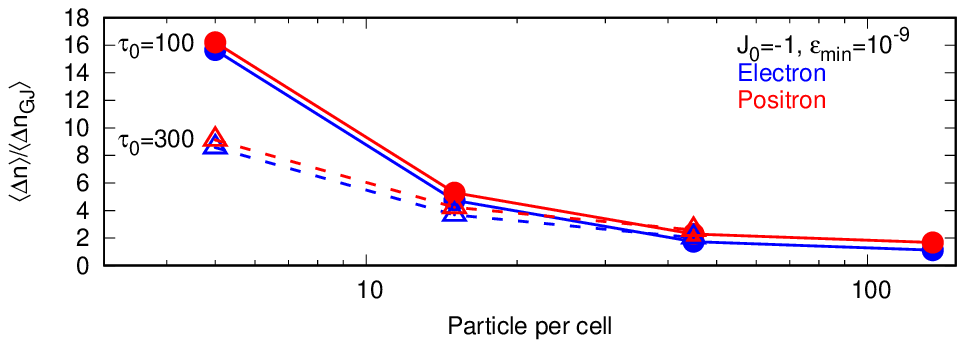}
   \caption{Time-averaged multiplicities of electrons (blue) and positrons (red) in the simulation box, as defined in Fig.\ref{fig:tau_dependence},  for different initial PPC.}
  \label{fig:ppc_dependence}
 \end{center}
\end{figure}

\subsection{Numerical effects}
\label{sec:numerical_effects}
As explained in \cite{2018A&A...616A.184L}, the skin depth depends on the multiplicity
and characteristic energy of pairs created by the oscillating electric field, and 
cannot be determined a priory.   Resolving the skin depth is crucial for avoiding 
artificial heating that can considerably affect the energy distributions of pairs and photons in the gap.  One therefore needs to perform convergence tests for each model. 
We verified that in all cases investigated the grid spacing was sufficiently small to
resolve the skin depth. 

Another important parameter is the initial PPC number. 
Fig. \ref{fig:ppc_dependence} shows the dependency of the time averaged multiplicities of electrons and positrons in the simulation box for two different choices of the physical parameters ($\tau_0=100$ and $\tau_0=300$, with the remaining parameters the same: $J_0=-1$, $\epsilon_{\min}=10^{-9}$, $p=2$).  As that figure indicates, a low initial PPC can lead to
artificially high multiplicities.  For the cases investigated here we find convergence for 
initial PPC of about 45.  Some other properties (e.g., the electric field dynamics) 
seem to be less sensitive to the PPC, although it can affect the prolong evolution, as described next.

As a comparison case we rerun the simulation described in \citet{2018A&A...616A.184L} (initial PPC of 5) for a longer time, and then repeated it with an initial PPC of 45 (model O). 
For the former case we find excellent agreement with the results of \citet{2018A&A...616A.184L} up to a time of $20 t_g/c$ (the final simulation time in \citealt{2018A&A...616A.184L}).  However, at longer times we observed a gradual opening of
the gap near the boundaries. 
In the second case, when the initial PPC was set at 45, the gap ultimately reached
a state of quasi-periodic oscillations at $\sim200 t_g/c$, like that seen in Fig. \ref{fig:j+1_tau100_snapshot},
and remained active over the whole simulation time.

\section{Discussion}
\label{sec:discussion}

\subsection{Pair creation beyond the boundary}

In our models, a fraction the pairs and photons escaping the outer boundary have energies in the range $\sim10-100$ TeV, high enough to induce further pair cascades. Consequently, it is conceivable that the pair multiplicity might further increase in the region outside the outer boundary. To check this, 
we estimate the number of created particles beyond the outer boundary, in the same way as in pulsar polar cap cascade models \citep{2015ApJ...810..144T,2017ApJ...837...76K}.
Since the number density of soft photons is significantly reduced beyond the outer radius of the soft photon emission zone, $R_{\rm s}$, it is sufficient to estimate the multiplicity up to this radius.
As seen from Fig. \ref{fig:j-1_lightcurve}, the IC gamma-ray luminosity is much higher than the particle kinetic luminosity. 
Thus, we only take into account pair production by the scattered gamma-ray photons. 
The multiplicity of newly created pairs beyond the outer boundary is given by
\begin{eqnarray}\label{eq:kappa_out}
\kappa_{\rm out}\sim2\frac{L_{\rm ic}}{4\pi r_{\rm out}^2m_{\rm e}c^3n_{\rm GJ}\epsilon_{\rm esc}},
\end{eqnarray}
where $\epsilon_{\rm esc}$ is the maximum energy of escaping photons that can reach the radius $R_{\rm s}$, defined implicitly through $\tau_{\gamma\gamma}(\epsilon_{\rm esc})\sim1$. 
Considering only the resonant reaction of pair creation, mono-energetic distribution of IC photons, and $4/\epsilon_{\rm esc}>\epsilon_{\min}$, the energy $\epsilon_{\rm esc}$ can be expressed as: 
\begin{eqnarray}\label{eq:epsilon_esc}
\epsilon_{\rm esc}\sim4\epsilon_{\min}^{-1}\left(\tau_0\frac{\sigma_{\gamma\gamma}}{\sigma_{\rm T}}\frac{R_{\rm s}}{r_g}\right)^{-1/p}.
\end{eqnarray}
For typical parameters, this energy is much higher than the energy of curvature radiation $\epsilon_{\rm c}$, so that pair production by the curvature photons can be safely neglected beyond the outer boundary.
Using Eqs. (\ref{eq:kappa_out}) and (\ref{eq:epsilon_esc}), the multiplicity is 
\begin{eqnarray}\label{eq:kappa_out2}
\kappa_{\rm out}\sim3\times10^{-2}&&\left(\frac{L_{\rm ic}}{10^{-5}L_{\rm BZ}}\right)B_3M_9\epsilon_{\min,-9} \nonumber \\
&\times&\left(100\tau_{0,2}\frac{R_{\rm s}}{10r_g}\right)^{1/p}. \\
 \nonumber
\end{eqnarray}
For example, for Model B and $R_{\rm s}=10r_g$, the last equation yields a multiplicity of newly created pairs beyond the outer boundary of $\kappa_{\rm out}\sim1$,
comparable to the average multiplicity in the simulation box, $\kappa\sim1$ (Fig. \ref{fig:tau_dependence}). 
This is consistent with the uniform multiplicity found in our simulations.
The above consideration is applicable to the $J_0<0$ current cases. However, if the gap is located within $\lesssim R_{\rm s}$, this estimate might also be applicable for the return current cases ($J_0>0$).
In the cases of a hard slope $p$, hard photon energy $\epsilon_{\min}$ and/or low escape energy $\epsilon_{\rm esc}<\epsilon_c$, pair creation beyond the outer boundary would significantly contribute to the total multiplicity.

\subsection{Comparison with other works}

Although the curvature radiation is neglected in other works \citep{2019PhRvL.122c5101P,2020ApJ...895..121C,2020PhRvL.124n5101C}, 
the luminosity could approach the BZ power in some Models shown in Section \ref{sec:results}. 
From the rough estimate, the curvature radiation dominates the total radiation reaction force if the Lorentz factor is larger than the value, 
\begin{eqnarray}
\gamma_{\rm eq}\sim\left\{ \begin{array}{ll}
1\times10^9\tau_{0}^{1/2}M_9^{1/2}\epsilon_{\min,-9}^{1/2} & ~(\gamma\epsilon_{\min}\lesssim1) \\
1\times10^9\tau_{0}^{1/3}M_9^{1/3} & ~(\gamma\epsilon_{\min}\gtrsim1), \\
\end{array} \right.
\end{eqnarray}
where we use $P_{\rm ic}=(4/3)\sigma_{\rm ic}u_{\rm s}\gamma^2c\sim(4/3)\epsilon_{\min}m_{\rm e}c^3\tau_{\rm ic}\gamma^2/r_g$ as the power of IC scattering, Eq. (\ref{sec3:tau_ic}) as optical depth $\tau_{\rm ic}$, and $R_{\rm c}\sim r_g$. 
The Lorentz factor can reach $\sim eE_{\parallel}l/m_{\rm e}c^2\sim10^{14}B_3M_9(E_{\parallel}/B)(l/r_g)$ in the absence of the reaction force.  
The radiative cooling via curvature radiation can increase the gap width and the resultant total dissipation energy as long as the curvature photons do not significantly contribute to the pair creation. 
For example, the total dissipation rates in Models B and D, where $L_{\rm cur}\lesssim L_{\rm ic}$, are comparable. 
However, the dissipation rate in Model C, for which $L_{\rm cur}>L_{\rm ic}$, is much higher than those in Models B and D. 
The condition $eE_{\parallel}l/m_{\rm e}c^2>\gamma_{\rm eq}$ is required for bright TeV flare in AGNs. 

Our model neglects the effects of the light surfaces. The 1D GRPIC model of \citet{2020ApJ...895..121C} includes the effects of the two light surfaces. Their simulation results show a quasi-periodic gap opening and screening at the null charge surface (their figure 1), very similar to our results. Thus, for 1D model, the existence of the light surfaces would not significantly affect the quasi-periodic gap activity. On the other hand, the results of 2D GRPIC simulations performed by \citet{2020PhRvL.124n5101C} show that the oscillating narrow gap opens near the inner light surface at their high optical depth case. At present it is difficult to compare 1D and 2D results because of the limitations introduced by the rescaling of 2D simulations, and the use of a very different photon spectra. Although a possible cause of the difference may be the difference of the main particle creation site (inner region of the inner light surface in 2D and between two light surfaces in 1D), such a comparison requires a more systematic investigation.

\subsection{Applications to M87}

The observed luminosity above 300 GeV is $L_{\rm TeV}\sim5\times10^{40}$ erg s$^{-1}$ in the quiescent state of M87 \citep{2020MNRAS.492.5354M}. The average jet power is estimated to be $\sim10^{43}$ erg s$^{-1}$ \citep[e.g., ][]{2006MNRAS.370..981S,2019ApJ...875L...5E}, which is consistent with the BZ luminosity for $M\sim6\times10^9M_{\odot}$, $B_{\rm H}\sim100$ G \citep{2015ApJ...803...30K,2019ApJ...875L...5E}, assuming $a_{\ast}\sim0.9$. Thus, the luminosity ratio in M87 is $L_{\rm TeV}/L_{\rm BZ}\sim10^{-4}-10^{-3}$. The spectral index and the cutoff energy of the disk emission inferred from observations (if indeed originating from an accretion disk) are $p\sim1.2$, and $\epsilon_{\min}\sim10^{-9}-10^{-8}$ which corresponds to $\approx0.5-5$ meV \citep{2016MNRAS.457.3801P}, respectively. If the observed photons at $\epsilon_{\min}$ are coming from near the horizon, the optical depth is $\tau_0\lesssim10^3$ \citep{2011ApJ...730..123L}. 

Our results indicate that the TeV luminosity is dominated by curvature radiation. In the case of $L_{\rm cur}/L_{\rm BZ}\sim10^{-3}$, the characteristic energy of curvature photons is $\epsilon_{\rm c}m_{\rm e}c^2\sim0.1-1$ TeV (Eq. \ref{eq:epsilon_c}). 
If the optical depth is $\tau_0\lesssim100$, the curvature luminosity in our model is consistent with the observations. Rapid variability ($\delta t\sim r_g/c$) of strong flares in the past decade \citep{2006Sci...314.1424A,2012ApJ...746..151A} is also consistent with  curvature radiation from the narrow gap. In the large optical depth case, the observed TeV emission may be explained by IC emission with $\epsilon_{\rm esc}m_{\rm e}c^2\sim1$ TeV. However, the cascade region should likely extend to the radius of the soft photon emission region, so that the variability timescale would be longer than that of the curvature radiation. Future observations by the Event Horizon Telescope will provide the parameters of the disk emission within $\sim10-100 r_g$ \citep[e.g., ][]{2019MNRAS.486.2873C,2019Galax...8....1H}, which may allow to distinguish between this model and its alternatives (e.g., synchrotron self-Compton model \citep{2020MNRAS.492.5354M}, and hadronic emission model \citep{2020arXiv200313173K}).

\section{Conclusions}
\label{sec:conclusion}

We conducted a comprehensive investigation of the dynamics of 1D gaps in a starved 
magnetosphere of a Kerr BH, using GRPIC simulations similar to those reported previously
in \cite{2018A&A...616A.184L} and \cite{2020ApJ...895..121C}, but for a broader range of spectral properties of the ambient radiation field, considerably longer run times, and different initial PPC and resolution.  Our main conclusions are:

(i) Sufficiently long run times (hundreds $r_g/c$) are required to allow the system to loose memory of the initial state and converge to its final state.  Under conditions 
for which the gap activity is maintained (roughly $\tau_0\gtrsim10$, depending on the 
spectral shape of the target spectrum), the system ultimately approaches a state
of quasi-periodic oscillations.  Otherwise, the vacuum state is ultimately restored and
the activity is switched off. 

(ii) The overall behaviour depends of the sign of the global magnetospheric current.  For
currents with the same sign of GJ current a quasi-periodic gap reopens near the null charge surface, as also 
found by \cite{2020ApJ...895..121C}. For the return currents it reopens near the 
boundaries of the simulation box.  In the latter case it is unclear whether this is an
artifact of the finite extent of the simulation domain, that prevents further pair production 
in the region outside the simulation box, as would occur in reality.

(iii) The amplitude of oscillations and the resultant luminosity of TeV emission 
depend sensitively on the spectral shape of the ambient radiation field, owing to
a combination of Klein-Nishina and threshold effects, as explained in Section \ref{sec:spectrum_dependence}.  The luminosity of IC TeV photons varies by no more than
an order of magnitude, and is around $10^{-5}L_{\rm BZ}$, as found in \cite{2018A&A...616A.184L}.
However, the curvature luminosity, that reflects the amplitude of oscillations 
of the quasi-periodic electric field can be as high as $10^{-2} L_{\rm BZ}$ (see Figs. \ref{fig:j-1_lightcurve} and \ref{fig:fig:j-1_tau100_epsilon-8_lightcurve}).  The 
sensitivity of the TeV luminosity to moderate changes of the soft 
spectrum of disk emission can lead to occasional strong flares, like those seen in M87.

(iv)  Too small initial PPC leads to artificially high pair multiplicity that 
might affect the solution.  For the models we tested, we found convergence for 
initial PPC of about 50.

\acknowledgments
We are grateful to the anonymous referee for constructive comments.
SK thank Shinpei Shibata for fruitful discussion. 
AL thank Benoit Cerutti for enlightening discussions.
Numerical computations were performed on Cray XC50 at cfca of National Astronomical Observatory of Japan, 
and on Cray XC40 at Yukawa Institute for Theoretical Physics, Kyoto University. 
This work was supported by JSPS Grants-in-Aid for Scientific Research 18H01245(K.T., S.K.), 18H01246, and 19K14712 (S.K.). 

\bibliographystyle{apj_8}
\bibliography{ref}

\appendix

\section{Space-time}

The background space-time is described by the Kerr metric given in Boyer-Lindquist coordinates with the notation 
\begin{eqnarray}\label{eq:coordinate}
ds^2=-\alpha^2dt^2+g_{\varphi\varphi}(d\varphi-\omega dt)^2+g_{rr}dr^2+g_{\theta\theta}d\theta^2
\end{eqnarray}
where
\begin{eqnarray}\label{eq:metric}
g_{rr}=\frac{\Sigma}{\Delta};~g_{\theta\theta}=\Sigma;~g_{\varphi\varphi}=\frac{A}{\Sigma}\sin^2\theta;
\alpha^2=\frac{\Sigma\Delta}{A};~\omega=\frac{2ar_gr}{A},
\end{eqnarray}
with
\begin{eqnarray}\label{eq:metric2}
\Delta&=&r^2+a^2-2r_gr;~\Sigma=r^2+a^2\cos^2\theta; \\
&A&=(r^2+a^2)^2-a^2\Delta\sin^2\theta, \nonumber
\end{eqnarray}
where $r_g=GM/c^2=1.5\times10^{14}M_9$ cm is the gravitational radius and $M_9=M/10^9M_{\odot}$ the BH mass.
The angular velocity of the BH is defined as $\omega(r=r_{\rm H})=a_{\ast}/(2r_{\rm H})$,
where $a_{\ast}=a/r_g$ is the dimensionless spin parameter and $r_{\rm H}=r_{\rm g}+\sqrt{r^2_{\rm g}-a^2}$ is the radius of horizon.
To avoid the singularity on the horizon, we use the tortoise coordinate $\xi$, defined by $d\xi= r_g^2 dr/\Delta$.  It is related to $r$ through:
\begin{equation}
\xi(r)=\frac{r_g}{r_+-r_-}\ln\left( \frac{r-r_+}{r-r_-} \right),
\end{equation}
with $r_\pm=1\pm \sqrt{1-a_{\ast}^2}$.  Note that $\xi\rightarrow-\infty$ as $r\rightarrow r_H=r_+$, and $\xi \rightarrow 0$ as $r\rightarrow\infty$.

\end{document}